# Single-shot 3D photoacoustic computed tomography with a densely packed array for transcranial functional imaging

Rui Cao[1#], Yilin Luo[1,2#], Jinhua Xu[1,2], Xiaofei Luo[1], Ku Geng[1], Yousuf Aborahama[1], Manxiu Cui[1,2], Samuel Davis[1], Shuai Na[1], Xin Tong[1,2], Cindy Liu[1,2], Karteek Sastry[1,2], Konstantin Maslov[1], Peng Hu[1], Yide Zhang[1], Li Lin[1], Yang Zhang[1], Lihong V. Wang[1,2*]

[1]Caltech Optical Imaging Laboratory, Andrew and Peggy Cherng Department of Medical Engineering, California Institute of Technology, Pasadena, CA, USA

[2]Caltech Optical Imaging Laboratory, Department of Electrical Engineering, California Institute of Technology, Pasadena, CA, USA

[#] These authors contributed equally: Rui Cao and Yilin Luo.

[*] Corresponding author: Lihong V. Wang. Email: lvw@caltech.edu

**Abstract:** Photoacoustic computed tomography (PACT) is emerging as a new technique for functional brain imaging, primarily due to its capabilities in label-free hemodynamic imaging. Despite its potential, the transcranial application of PACT has encountered hurdles, such as acoustic attenuations and distortions by the skull and limited light penetration through the skull. To overcome these challenges, we have engineered a PACT system that features a densely packed hemispherical ultrasonic transducer array with 3072 channels, operating at a central frequency of 1 MHz. This system allows for single-shot 3D imaging at a rate equal to the laser repetition rate, such as 20 Hz. We have achieved a single-shot light penetration depth of approximately 9 cm in chicken breast tissue utilizing a 750 nm laser (withstanding 3295-fold light attenuation and still retaining an SNR of 74) and successfully performed transcranial imaging through an *ex vivo* human skull using a 1064 nm laser. Moreover, we have proven the capacity of our system to perform single-shot 3D PACT imaging in both tissue phantoms and human subjects. These results suggest that our PACT system is poised to unlock potential for real-time, *in vivo* transcranial functional imaging in humans.

## Introduction

Functional brain imaging holds great importance, granting us access to the intricate workings of the brain—an essential organ that directs our cognition, emotions, and behaviors. These imaging modalities enable us to visualize brain activity, thereby aiding in the diagnosis and treatment of neurological disorders, advancing our understanding of cognitive processes, and expanding our knowledge of brain function. The current arsenal of functional brain imaging technologies, including functional magnetic resonance imaging (fMRI), electroencephalography (EEG), positron emission tomography (PET), functional near-infrared spectroscopy (fNIRS), and functional ultrasound imaging, has enriched our understanding of the brain. Nevertheless, they come with limitations concerning temporal, spatial resolution, and radioactivity. fMRI suffers from poor temporal resolution[1], whereas EEG's limitation lies in its spatial resolution[2]. While PET facilitates metabolic brain imaging, it necessitates the use of radioactive agents[3]. fNIRS faces constraints due to its suboptimal spatial resolution[3]. Transcranial functional ultrasound imaging, usually relying on contrast agents to visualize tiny blood vessels and flow[4], faces hurdles in imaging brain oxygenation and metabolism.

Emerging into the landscape of brain imaging, photoacoustic computed tomography (PACT) combines the strengths of light and sound, providing high-resolution images of tissues and organs via light-induced ultrasound signals[5]. It presents promising prospects for transcranial imaging. By harnessing the optical molecular specificity of oxy-hemoglobin and deoxy-hemoglobin, PACT enables precision mapping of cerebral blood oxygenation and functional assessment of the brain. Unlike pure ultrasound imaging, PACT undergoes one-way ultrasound transmission through the skull, reducing acoustical distortions and promising visualization of cortical brain functions. Moreover, PACT operates without magnets in a non-radioactive manner, enhancing its accessibility, cost-effectiveness, and mitigating risks for patients. The label-free nature of PACT circumvents the need for contrast agents, enabling real-time imaging of brain function without exogenous substances. These unique features position PACT as a strong candidate for transcranial imaging, paving the way for new pathways in studying and understanding brain functions.

Though PACT has shown promise in functional brain studies in small and large animals[6–9], its application in human transcranial imaging faces challenges due to the thickness of the skull and the acoustic attenuation and wavefront aberration encountered in human cranial bone. While initial investigations into transcranial PACT imaging through human skull bone in phantom models and *ex vivo* skulls are promising[10,11], *in vivo* demonstration of PACT through the human skull remains challenging[12]. However, significant progress has been made in patients having undergone hemicraniectomy[13], where part of the skull is absent. The results not only correlated well with fMRI findings but also offered a more detailed understanding of blood vessel structures, underscoring PACT's potential in functional brain imaging. Yet, challenges with the current PACT system for fast transcranial functional brain imaging persist, which requires a rotating mechanism for 3D imaging due to the sparse distribution of the transducer array[13,14]. Also, our existing PACT system was originally optimized at a central frequency of 2.25 MHz for breast imaging, leading to substantial acoustical attenuations by the human skull and weak signals for transcranial PACT brain imaging. These issues highlight the need for ongoing innovation and improvements in PACT system design and optimization to further advance transcranial functional brain PACT imaging.

Addressing the challenge of photoacoustic transcranial brain imaging, we have designed a state-of-the-art PACT system. This system leverages a densely packed hemispherical transducer array, incorporating 3072 channels and operating at an optimal central frequency of 1 MHz for transcranial imaging in humans. The array's hemispherical layout allows single-shot 3D imaging by avoiding the need for array rotation. This unique characteristic propels it as an excellent candidate for rapid functional brain imaging, offering enhanced spatial and temporal resolutions. To the best of our knowledge, this configuration of the transducer array presents the highest number of parallel channels ever used in PACT systems. Augmenting its capabilities further, our PACT system has the solid angle that is 206 times larger than our previous design[13–16], ensuring a substantial increase in sensitivity for transcranial photoacoustic imaging. From our system characterization and results obtained from phantom imaging and *in vivo* imaging, our PACT system has demonstrated advantageous prospects for achieving transcranial PACT imaging in humans.

**PACT with a densely packed 3072-element hemispherical transducer array (3K-PACT)**

To achieve high acoustical detection sensitivity and spatiotemporal resolution, we have custom-built a densely packed 3072-element hemispherical ultrasonic transducer array. This hemispherical array, with a 150 mm radius, performs at a central frequency of 1 MHz and exhibits a one-way –6dB bandwidth of ~81%. A 94% of the hemisphere's surface, excluding a central hole designated for light illumination and the mounting frames, is covered by ultrasonic transducers, thereby maximizing acoustical detection sensitivity. This design enables single-shot 3D isotropic imaging without the need for rotation.

To minimize noise and guarantee a high signal-to-noise ratio, we directly linked the array connectors to 24 printed circuit boards (PCBs) equipped with 128-channel preamplifiers, which are also directly connected to 12 data acquisition modules (DAQs) with 256 channels. We synchronized the 12 DAQs with a hub module via HDMI cables. The acquired photoacoustic signals were then digitized and directly streamed to the solid-state drive via 12 USB 3.0 ports, allowing high-speed data transfer (> 480 MB/s).

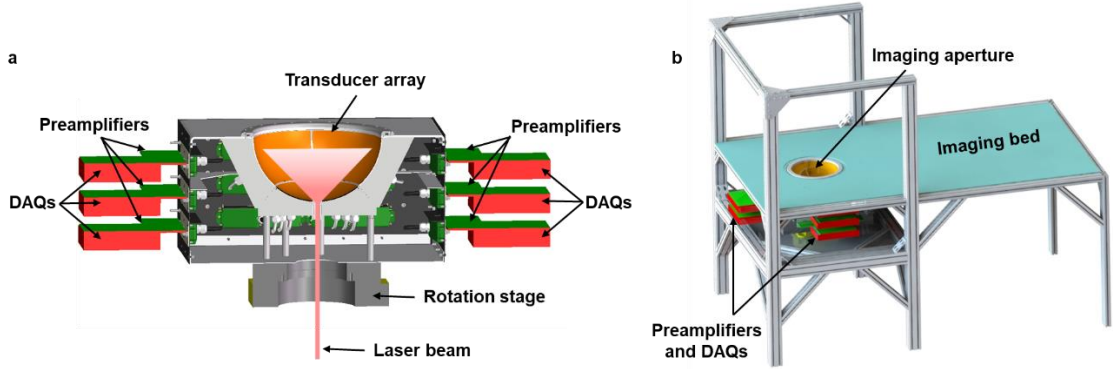

**Figure 1. Representations of the 3K-PACT**. **a** Cut-away view of the 3K-PACT imager. **b** Perspective view of the 3K-PACT system. DAQ, data acquisition module.

To acquire 3D PACT images, we steer the laser beams through the hole at the base of the hemisphere. The beams are subsequently homogenized and expanded using a diffuser prior to illuminating the target object (Fig. 1a). For functional PACT imaging, we deploy two synchronized lasers with wavelengths of 750 nm and 1064 nm. To ensure a comfortable human imaging experience, we have integrated our 3K-PACT system into a specially designed imaging bed where participants can lie down comfortably (Fig. 1b). The imaging aperture is filled with water to effectively couple the photoacoustic signals. We have carefully regulated the surface laser fluences (i.e., radiant exposures) on the imaging plane to adhere to the American National Standards Institute (ANSI) safety limits for laser exposure — 100 mJ/cm$^2$ at 1064 nm and 25 mJ/cm$^2$ at 750 nm with the laser repetition rate of 10 Hz[17]. To monitor pulse-by-pulse energy fluctuations, we have positioned two photodiodes near the laser output. These fluctuations are recorded using a multifunction I/O device, which also synchronizes the two lasers, rotation motor, photodiodes, and DAQs via custom software complemented with a MATLAB 2022-based graphical user interface (GUI). This user-friendly GUI enables synchronization of multiple instruments, recording of photodetector data, real-time display of raw photoacoustic signals, and 3D image reconstruction. It streams data directly from 12 DAQs to the computer memory, bypassing the time-consuming process of data download and format conversion, which has traditionally been a bottleneck due to extensive data acquisition processes. Our purpose-built software also facilitates pre-save data processing, including averaging and filtering, providing real-time 3D reconstruction for feedback during human imaging experiments.

**Geometric calibration of the 3K-PACT system**

Manufacturing in the transducer array may result in deviations of the actual locations of the 3072 transducer elements from their designed positions. Using these designed element coordinates directly for image reconstruction could compromise the quality of the resulting image. As a countermeasure, a geometric calibration can identify the true positions of the elements. By replacing the ideal coordinates with these accurately calibrated ones, we can noticeably enhance the image quality. The calibration process involves recording photoacoustic signals from a point emitter that moves in a predefined, controlled pattern. The photoacoustic point source was generated using a multimode fiber, with its tip coated in an epoxy mixed with carbon powder. The strong absorption properties of the carbon cause it to emit photoacoustic signals when the laser is guided through the fiber to the tip. We employed a 3D motorized stage to accurately control the relative positions of the point source. Data were collected from 216 distinct positions, creating a 3-dimensional grid of 6×6×6 with a 7 mm step size in the *x* and *y* axes and a 5 mm step size along the *z*-axis. The distance from the point source to each of the elements was computed by measuring the signal's time delay and multiplying it by the speed of sound, which we gauged through water temperature measurement. Additional details regarding the geometric calibration method can be found in our previous study[18].

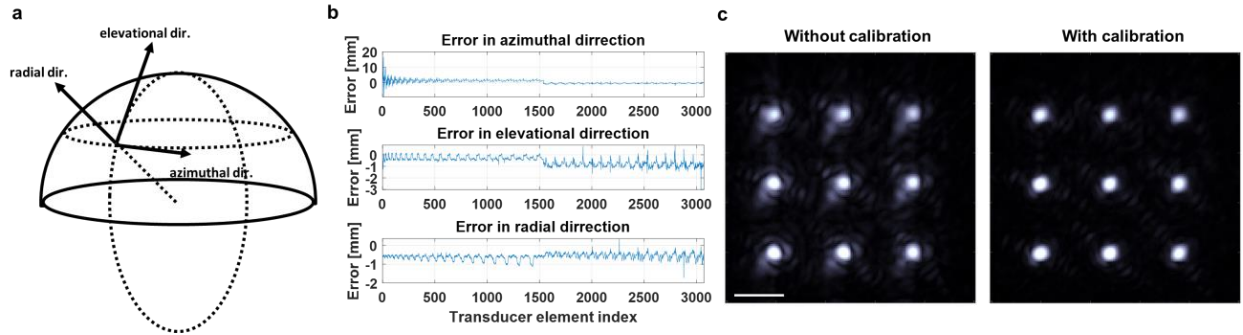

**Figure 2. Geometric calibration of the 3K-PACT array. a** Representation of three directions in the array geometric calibration. **b** Errors segmentation into azimuthal, elevational, and radial directions. **c** Reconstruction of a scanned point source pattern using designed element positions and calibrated element positions. Scale bar, 5 mm.

Using these calculated distances and point source scanning locations, we managed to ascertain the coordinates of the elements. The manufacturing error was quantified by first applying a rigid transformation, encompassing rotation and translation, to minimize the discrepancies between the ideal and measured element positions. Upon completing the fitting, the errors were segmented into three directions, as illustrated in Fig. 2a. The differing patterns observed between the first and second halves of the elements suggest a global drift between the upper and lower panels of the hemisphere bowl (Fig. 2b). In Fig. 2c, we compared image reconstruction results using both the designed and calibrated geometry. The noticeably suppressed background artifacts following calibration provide evidence of effective implementation, leading to enhanced image quality.

**Model-based iterative reconstruction for the 3K-PACT system**

Several techniques are available for image reconstruction in PACT, with time-reversal-based techniques like universal back-projection (UBP) and model-based iterative techniques being among the most prevalent. The former boasts advantages like straightforward implementation and high computational efficiency. In contrast, the latter forgoes some of these benefits in favor of greater flexibility, accommodating a multitude of different effects such as each transducer's spatial impulse response (SIR). Moreover, these iterative methods can offer joint reconstruction and more robust tuning.

For the 3K-PACT, the implemented model-based technique operates in the frequency domain and employs the free space propagation Green's function for the Helmholtz equation to depict the point spread function (PSF). When SIR is incorporated, the PSF is integrated on each transducer surface to estimate its SIR. After constructing the system matrix that correlates the image (sources) to the transducers' measurements (at each frequency), the reconstruction can be expressed as an optimization problem. The objective function includes a weighted $L^2$-norm for the disparity between the modeled field and the measured one. This problem is then transformed into a linear system of equations solved using the generalized minimum residuals (GMRES) method. Additional constraints and regularizers, such as positivity and total variations (TV), can be incorporated into the optimization problem. The optimization problems addressed here are convex, and the proximal gradient descent method can be effectively used to resolve them.

Figure 3a presents a comparative analysis of the reconstructed images of a scanned 6×6 point source pattern, employing both UBP and model-based iterative techniques. The imaging process did not involve array rotation to accurately represent single-shot PACT images. Close to the center of the ideal field of view, UBP manages to reconstruct the point sources satisfactorily, but sidelobe-like artifacts around the points are still identifiable (Fig. 3b). These artifacts might have resulted from the nonuniform sampling of 3K-PACT. In contrast, the model-based reconstruction method greatly mitigates these sidelobe-like artifacts, resulting in a clearer background. At the edge of the PACT field of view, both UBP and the model-based iterative reconstruction techniques encounter more pronounced blurring. However, the model-based reconstruction technique that incorporates the SIR restored the distorted points to their anticipated shapes (Fig. 3c). This comparative analysis clearly demonstrates that the model-based reconstruction approach with SIR offers superior performance in terms of artifact suppression and distortion resistance over traditional UBP. This advantage is particularly beneficial for the 3K-PACT system, given its nonuniform sampling.

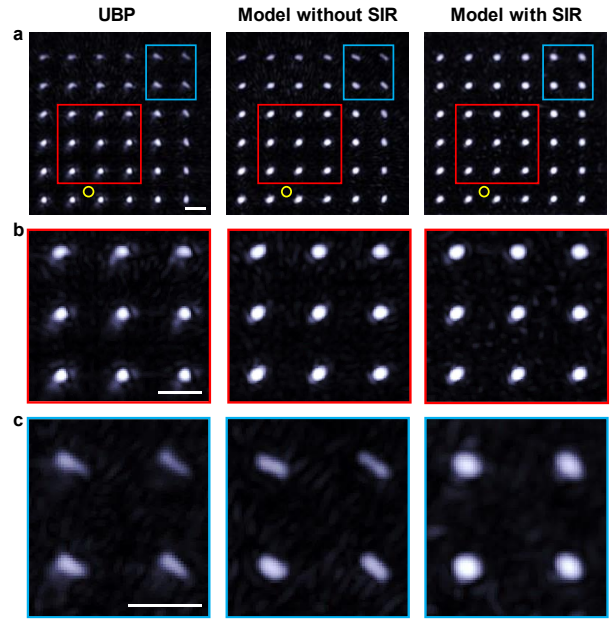

**Figure 3. Comparison between different reconstruction methods**. **a** *XY* projection images of 3D PACT reconstruction of a 6×6 point source pattern by UBP, model-based iterative reconstruction without SIR and with SIR. The yellow circles indicate the transducer array center position. **b** Close-up images near the center to show reconstruction artifacts. **c** Close-up images near the edge to show reconstruction distortions. Scale bars, 5 mm.

**PACT imaging with deep light penetration**

The ultra-sensitive 3K-PACT system with its 3072 elements, offers a practical platform for deep-tissue photoacoustic imaging in human subjects. We evaluated its performance through phantom experiments involving chicken breast tissue of varying thicknesses. This tissue was chosen to simulate the light attenuation characteristics of human tissues at wavelengths of 1064 nm and 750 nm. As illustrated in Fig. 4a, laser beams were oriented from the top of the transducer array, traveling through the chicken breast tissue before illuminating a black wire target with a beam diameter of around 5 cm. We confirmed that the surface laser fluence (i.e., radiant exposure) remained below the ANSI safety limits. Moreover, we took precautions to completely cover the top space of the PACT array, preventing any stray light from reaching the object.

We achieved single-shot imaging depths of 9 cm in chicken breast tissue at 750 nm wavelength and 6 cm at 1064 nm wavelength (Fig. 4b). The images were reconstructed by the forward model iterative methods. SNRs were evaluated at different penetration depths, revealing an exponential decay pattern (i.e., linear decay on a logarithmic scale), an indication that no noticeable stray light traversed a shorter beam path to the object. At 1064 nm, the photoacoustic signal SNR showed a decline with a slope of 1.1/cm, while 750 nm demonstrated a slower attenuation with a slope of 0.9/cm. With ~9 cm of chicken breast tissue that attenuated light by 3295-fold at 750 nm, we observed an SNR of 74 in the reconstructed single-shot photoacoustic image. This experiment demonstrates that the 3K-PACT is capable of single-shot deep-tissue photoacoustic imaging.

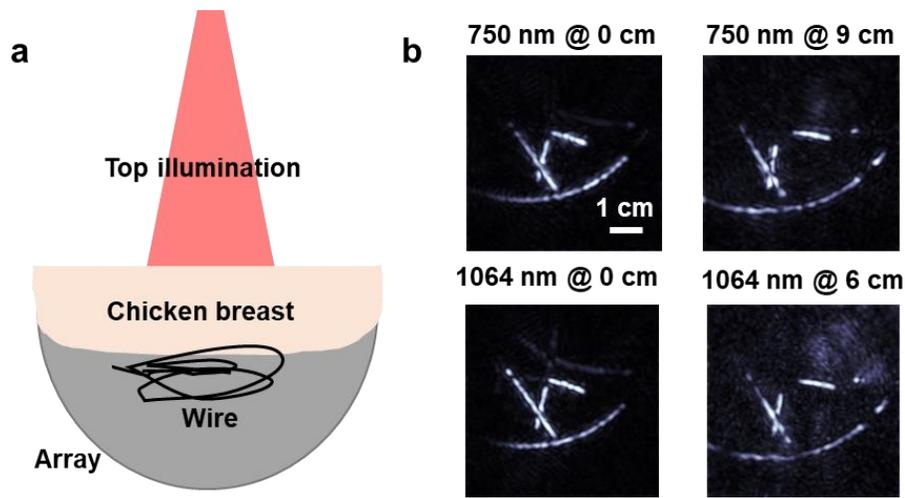

**Figure 4. Single-shot PACT imaging through *ex vivo* tissue of varying thicknesses**. **a** Schematic of the experiments. **b** Single-shot PACT images acquired without and with chicken breast tissues of different thicknesses with 750 nm and 1064 nm lasers, respectively.

**Feasibility study of transcranial 3D-PACT imaging**

To demonstrate the feasibility of transcranial PACT imaging, we imaged *ex vivo* a human skull obtained from a donor (Fig. 5a). The object consisted of a wire placed inside the human skull, simulating the transcranial PACT imaging of cortical vessels in human brains. The 1064 nm laser light illuminated the skull from the bottom hole on the transducer array. The skull and the optically absorbed wire were placed in proximity to the center of our hemisphere array, ensuring that they fell within the desired field of view (Fig. 5b). Despite the acoustic attenuation and distortion induced by the skull, the 3D structure of the wire could still be identified in the transcranial PACT images, although with some blurring (Fig. 5c). This demonstration highlights the potential of visualizing and studying cortical vessels in human brains non-invasively, through the skull. In addition, we demonstrated *in vivo* 3D PACT imaging of a human hand, from which vessels at different depths can be identified (Fig. 5d).

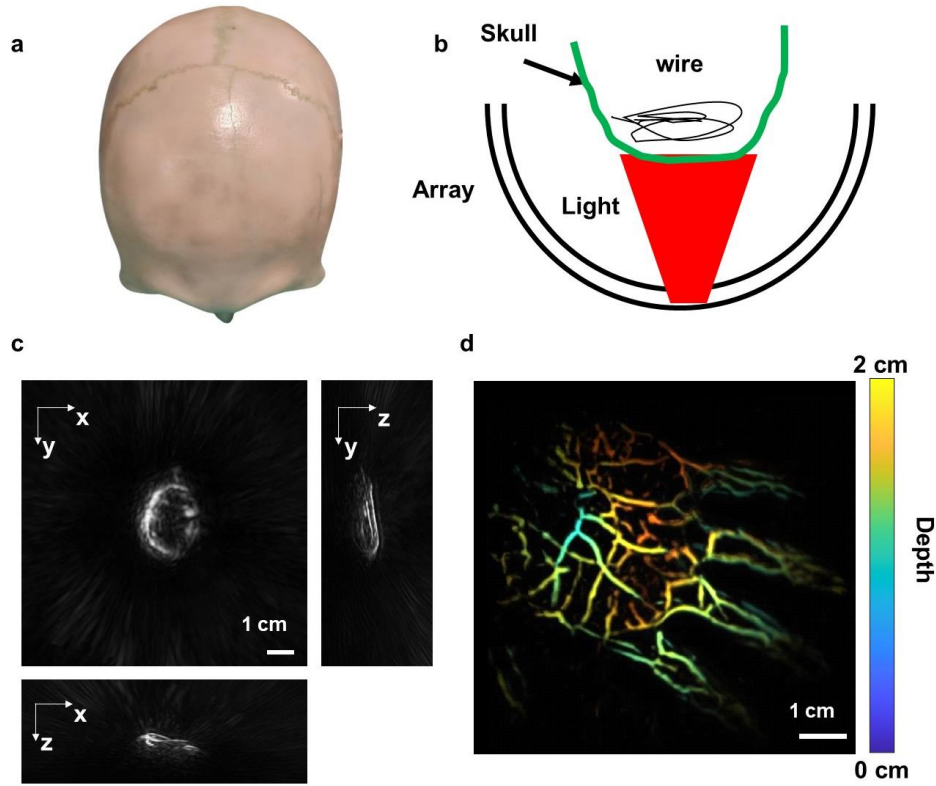

**Figure 5. Demonstration of transcranial 3D PACT imaging and fast *in vivo* imaging. a** Photo of human skull from a donor used in the phantom study. **b** Schematic of the *ex vivo* transcranial imaging experiment. **c** Transcranial images of the wire. **d** Depth-encoded *in vivo* image of a human hand.

**Discussion and summary**

Our 3K-PACT system has demonstrated its performance in deep tissue imaging, setting a new standard in spatiotemporal resolution and sensitivity. The system's unique design enables single-shot 3D PACT imaging without necessitating rotation, paving the way for rapid functional brain imaging. By geometrically calibrating the 3072-element ultrasonic array, we have notably optimized the system's performance, resulting in a marked improvement in image quality. We have employed a model-based iterative method to effectively mitigate background artifacts. Coupling this approach with incorporated SIRs has allowed for further enhancement of image quality, which minimizes reconstruction distortion at the edge of the ideal field of view and boosts the contrast-to-noise ratio. Our choice of 1 MHz frequency, specifically optimized for transcranial imaging, is expected to yield a minimum tenfold increase in transcranial ultrasound transmittance compared to our previous 2.25 MHz system[19]. With a solid angle approximately 206 times larger than that of our previous iteration due to the dense packing, the 3K-PACT improves sensitivity and shows the potential for transcranial functional photoacoustic imaging.

Our results demonstrate that the 3K-PACT system can convincingly image phantoms up to a light penetration depth of 9 cm through chicken breast using a wavelength of 750 nm with a single laser shot. The reconstructed images at this depth still maintain a relatively high contrast-to-noise ratio, suggesting the potential for imaging beyond the 9 cm penetration depth *ex vivo*. Employing approximately 9 cm of chicken breast tissue, which attenuated light by a factor of 3295-fold at 750 nm, we recorded an SNR of 74 (i.e., 37 dB) in the reconstructed single-shot photoacoustic image. Similarly, with roughly 6 cm of chicken breast tissue that attenuated light by 735-fold at 1064 nm, we noted an SNR of 49 (i.e., 34 dB) in the corresponding single-shot photoacoustic image. A previous study demonstrated PACT imaging of surfactant-stripped micelles through 12 cm of chicken breast tissue at 1064 nm[20]; however, their results exhibited a lower SNR of 24 dB even with 100-times averaging, equivalent to about 4 dB SNR in single-shot PACT images. It is also worth noting that the previous study showed much less light attenuation despite the thicker chicken breast tissue, which was possibly due to differences in chicken breast tissue properties and preparations. Based on the SNR attenuation curve described in that study[20], the light attenuation of a 12-cm chicken breast they encountered was around 19-fold at 1064 nm, which is markedly lower than the light attenuation we experienced with either 9-cm chicken breast tissue at 750 nm (3295-fold) or 6-cm of chicken breast tissue at 1064 nm (735-fold). We validated the feasibility of transcranial imaging utilizing an optically absorbing object enclosed with a real *ex vivo* human skull. We plan to proceed with transcranial human brain imaging using the 3K-PACT.

## Data availability

The data that support the findings of this study are provided within the paper.

## Code availability

The reconstruction algorithm is fully described in the manuscript. The reconstruction code is not publicly available because it is proprietary and is used in licensed technologies.

## Contributions

L.V.W. and R.C. designed the study. R.C., Y.L., J.X., G.K., X.L, and X.T. built and optimized the 3K-PACT system hardware. L.V.W., S.D., S.N., K.M., L.L., and Yang.Z. designed the array. R.C. and Y.L. performed the PACT experiments. Y.A., M.C., Y.L., and S.D. developed the 3D PACT reconstruction algorithm. R.C., Y.L., and Yide.Z. developed a control software. R.C., Y.L., M.C., and K.S. conducted the geometric calibration. R.C., Y.L., M.C., Y.A., and C.L. analyzed and interpreted the data. All authors contribute to the manuscript writing. L.V.W. supervised the study.

## Acknowledgments

This work was sponsored by the US National Institutes of Health (NIH) grants U01 EB029823 (BRAIN Initiative) and R35 CA220436 (Outstanding Investigator Award).

## Competing interests

L.V.W. has a financial interest in Microphotoacoustics Inc., CalPACT LLC, and Union Photoacoustic Technologies Ltd., which, however, did not support this work.